\documentclass[]{l4dc2026}

\title[]{The Impact of Scenario Representation in Scenario-based Optimization Certification}
\usepackage{times}

\newtheorem{thm}{Theorem}
\newtheorem{prop}{Proposition}

\newtheorem{cor}{Corollary}
\newtheorem{ass}{Assumption}

\theorembodyfont{\normalfont}
\newtheorem{defin}{Definition}

\theorembodyfont{\normalfont\itshape}

\author{%
 \Name{Stephen Crawford} \Email{stephen\_crawford@brown.edu}
 \AND
 \Name{Nora Ayanian} \Email{nora\_ayanian@brown.edu}\\
}

\begin{document}

\maketitle

\begin{abstract}
Scenario-based optimization provides finite-sample guarantees for decisions constructed from sampled uncertainty, but its certification procedures can be sensitive to how those samples are represented. We investigate this sensitivity under lossless scenario regrouping, which preserves the sampled information, feasible region, optimizer, and primitive violation probability. For an equal-allocation support-subsample certificate, we derive the distortion factor \(k^2(1-\epsilon)^{k-1}\), establish certification reversals between equivalent representations, and provide sufficient conditions for acceptance dominance under nested coarsening. We further apply nondegeneracy-free Wait-and-Judge theory to coordinatewise scenario envelopes, enabling stronger certification directly from the primitive observations. In paired scenario-MPC experiments, exact regrouping increases certification availability from 318 to 471 of 480 attempts without changing the samples, numerical proposals, or requiring any reduction in witness cardinality. Primitive Wait-and-Judge certification accepts the same 471 attempts without regrouping. These results demonstrate that apparent scenario sample-efficiency gains can arise from certificate accounting rather than additional information or reduced decision risk.
\end{abstract}


\section{Introduction}\label{sec:introduction}
Scenario-based optimization is a powerful technique which allows for chance-constrained optimization problems to be solved with a computationally tractable surrogate which offers favourable probabilistic certificates. In turn, scenario-based methods have been developed for a range of problems including power supply allocation (\cite{geng2019chanceconstrainedunitcommitmentscenario, ROALD201566}), portfolio optimization (\cite{Pagnoncelli2012RiskReturnTW}), and model predictive control (MPC) (\cite{de2021ScenarioBasedTraj,degroot2023SBMotionBounded, Shang2021SceMPC, Jekl2025SceVelPlanning, crawford2026distributiontransfersafehorizonmpcmode}). Across applications, a consistent limitation of scenario-based optimization is its sample inefficiency. This has precipitated a wide range of research which describe problem-specific (\cite{keutchayan2021problemdrivenscenarioclusteringstochastic, cordiano2024scenarioreductionguaranteesstochastic}) and general methods (\cite{henrion2009scenario, rujeerapaiboon2017scenarioreductionrevisitedfundamental, choi2024reducedsamplecomplexityscenariobased}) for reducing this complexity. 

Following \cite{Cal2006ScenarioApproach}, we review the original concept of the so called scenario-approach for chance-constrained optimization problems in order to demonstrate the sample complexity issue as well as motivate our investigation into scenario-based certificates themselves. Let $w\in\mathbb R^d$
be a deterministic
linear-objective vector. Consider a chance-constrained optimization problem, 
\begin{align}
\mathrm{CCP}(\epsilon):
\quad
\min_{\theta\in\Theta\subseteq\mathbb{R}^d}
&\ w^\top\theta
\nonumber\\
\text{subject to}\quad
&P\{f(\theta,\delta)\leq0\}\geq1-\epsilon,\label{eq:ccp}
\end{align}
where $f(\theta, \delta) \leq 0$ is a constraint evaluation function on decision $\theta \in \Theta$ and any uncertainty $\delta \in \Delta$ in uncertainty probability space $(\Delta, \mathcal{D}, P)$, and $\epsilon$ is a prescribed violation tolerance. This form typically assumes that $f(\theta, \delta)$ is convex in $\theta$ and is generally intractable. Scenario optimization replaces the
chance constraint with sampled constraints following $\omega_N
=
(\delta_1,\ldots,\delta_N)
\in \Delta^N \sim P^{\otimes N}$. These constraints are then enforced as \begin{align}\label{eq:scp}
\mathrm{SCP}_N:
\quad
\min_{\theta\in\Theta\subseteq\mathbb{R}^d}
&\ w^\top\theta
\nonumber\\
\text{subject to}\quad
&f(\theta,\delta_i)\leq0,
\quad i=1,\ldots,N.
\end{align} 

Given \eqref{eq:scp}, it is common to  want to form a finite-sample certificate on the violation probability. For convex scenario programs, classical scenario theory relates the violation probability of the random optimizer to a binomial-tail bound. For uncertainty
$\delta$, define its
feasible decision set $\Theta_\delta
:=
\{\theta\in\Theta:
f(\theta,\delta)\le0\}$. The violation probability
of a decision $\theta$ is then given as $V_P(\theta)
:=
P\{\delta\in\Delta:
\theta\notin\Theta_\delta\}$. Under these definitions and the standard convexity and solution-selection assumptions, \cite{Campi2008} states that \begin{equation}
\label{eq:binom_cert}
P^{\otimes N}
\left\{
V_P(\theta_N^\star)>\epsilon
\right\}
\leq
\sum_{i=0}^{d-1}
\binom{N}{i}
\epsilon^i(1-\epsilon)^{N-i}.
\end{equation} where $\theta^\star_N$ is the solution to $\mathrm{SCP}_N$ and $d$ is the dimension of the decision $\theta$. This bound is exact for
fully supported problems, whose
solutions have $d$ support constraints
almost surely. However, for problems that are not fully
supported, the dimension-based
bound remains valid but is not necessarily tight. This motivates a posteriori
certification methods that account
for the realized support complexity, and  will play a role in our later theoretical analysis. 

As can be seen in \eqref{eq:binom_cert}, the sample complexity of the problem can grow prohibitive when we hope to restrict tsurrogatehe violation risk and/or the probabilistic certificate. To this point, scenario-reduction research investigates ways to either generate additional samples \cite{dupacova2003scenario, axelrod20} or maintain similar certification under fewer samples using additional assumptions or statistical tools  \cite{rujeerapaiboon2017scenarioreductionrevisitedfundamental, keutchayan2021problemdrivenscenarioclusteringstochastic, cordiano2024scenarioreductionguaranteesstochastic}. 

In lieu of extending these lines of work, we raise the adjacent question of whether commonly used finite-sample certificates are themselves invariant to how the same sampled information is represented. More specifically, we investigate whether finite-sample certificates are invariant under lossless changes to the atomic representation of the sampled observations. We first establish that exact grouping preserves the sampled feasible region and induces an exact transformation of primitive violation probability. We then construct a scenario program for which singleton and grouped representations contain identical information, return the same optimizer, and exhibit the same primitive violation law, yet receive different assessments under an elementary support-subsample certificate. For a scalar scenario program, we
quantify this discrepancy through
the elementary certificate-distortion
factor $R_\epsilon(k)$, defined as
the ratio of singleton to grouped
certificate factors when $k$
primitive observations are placed
in each exact block. We establish $R_\epsilon(k)
=
k^2(1-\epsilon)^{k-1}$. We additionally show sufficient conditions for certificate-acceptance dominance under nested coarsening and investigate how stronger, structure-aware certification can eliminate the observed behavioural differences. 

\section{Theoretical analysis}\label{sec:derivation}

Recalling the definitions provided in the introduction, we begin our theoretical analysis by describing the basic statistical framework of our problem. We first establish the statistical equivalence induced by exact grouping, then quantify the resulting elementary-certificate distortion and derive a coarsening dominance condition. Finally, we apply primitive Wait-and-Judge certification to coordinatewise envelopes. Proofs appear in the appendix.

\subsection{Scenario decisions and primitive risk}

Throughout our derivation, we assume the measurability of our statistics. Further, we hold the typical assumption that any decision returned by our generic optimizer satisfies the sampled constraints: 
\begin{ass}[Sampled feasibility]
\label{ass:feasible_family}

For every $m\ge0$, let
$A_m:\Delta^m\to\Theta$
be a measurable deterministic
decision map. For every $m\ge1$ and
$\delta_{1:m}\in\Delta^m$,
assume $A_m(\delta_{1:m})
\in
\bigcap_{i=1}^{m}\Theta_{\delta_i}$.\end{ass} 

\noindent We now introduce our first novel definition. 

\begin{defin}[Lossless scenario representation]
\label{def:lossless_representation}

Let $\mathcal U\subseteq\Delta^N$ be a
measurable set of primitive observation banks.
For a measurable representation space
$(\mathcal R,\mathcal S)$, equip the image
$R(\mathcal U)$ with the trace
$\sigma$-algebra. A measurable map
$R:\mathcal U\to\mathcal R$
is \emph{lossless} on $\mathcal U$
if a measurable decoder
$D:R(\mathcal U)\to\mathcal U$
satisfies $D\circ R=\operatorname{id}_{\mathcal U}$. Two representations $R_a$ and $R_b$
are \emph{losslessly equivalent}
if both are lossless on the same
$\mathcal U$. In that case, $T:=R_b\circ D_a:
R_a(\mathcal U)\to R_b(\mathcal U)$ is a measurable bijection with
measurable inverse $R_a\circ D_b$,
and $T(R_a(\omega_N))=R_b(\omega_N)$
for every $\omega_N\in\mathcal U$.
\end{defin}

\noindent Extending the notion of losslessly equivalent representations, we say that associated decision maps
$A_a$ and $A_b$ are
\emph{decision equivalent}
when $A_a(R_a(\omega))
=
A_b(R_b(\omega))
\quad\forall\omega\in\mathcal U$. Similarly, for representation-specific certificate
acceptance predicates $\chi_a$ and $\chi_b$,
we call the certification procedure
\emph{representation invariant}
over a declared class if $\chi_a(R_a(\omega_N);\epsilon,\beta)
=
\chi_b(R_b(\omega_N);\epsilon,\beta)$
for every admissible primitive bank
and each losslessly equivalent,
decision-equivalent pair in that class. Given these definitions, we are now able to to introduce our first basic construction in the form of an exact $k$-block representation. This form will be key to our later theorem on certificate behaviour.

\begin{defin}[Exact block representation]
\label{def:exact_k_block}

Let $k\in\mathbb N$ be a block size
dividing the primitive sample count,
written $k\mid N$, and define
$M_k=N/k$. Further define $Z_j^{(k)}
=
(\delta_{(j-1)k+1},\ldots,\delta_{jk}),
\; j=1,\ldots,M_k$. Then the the exact block-representation map $R_k:
\Delta^N\to(\Delta^k)^{M_k}$ is given by $R_k(\omega_N)
=
(Z_1^{(k)},\ldots,Z_{M_k}^{(k)})
=:\Omega_k$. It is easy to see that the original primitive bank is
recovered by concatenating the
ordered blocks, meaning $R_k$ is lossless. As an ordered tuple of
$k$ independent primitive observations,
we denote its probability law as $Q_k:=P^{\otimes k}$. Additionally, because the blocks use disjoint primitive
indices, the grouped bank contains
$M_k=N/k$ i.i.d. observations
under $Q_k$. Thus for $Z=(\delta_1,\ldots,\delta_k)$, let $\Theta_Z
=
\bigcap_{r=1}^{k}\Theta_{\delta_r}
$.
\end{defin}

\noindent Under our definition of exact block representations, the sampled feasible region is preserved:
\begin{equation}
\label{eq:feasible_region_identity}
\bigcap_{j=1}^{M_k}\Theta_{Z_j^{(k)}}
=
\bigcap_{i=1}^{N}\Theta_{\delta_i}.
\end{equation}
Despite this, we will shortly see that it does not guarantee identical acceptance decisions for a given certificate. We next relate primitive violation risk to its block counterpart in order to show that the associated risk is equivalent.

 \begin{prop}[Exact block-risk transformation]
\label{prop:block_risk_transform}

For any fixed decision $\theta\in\Theta$, define the block violation probability as $V_{Q_k}(\theta)
=
Q_k\{Z:\theta\notin\Theta_Z\}$. Then
\begin{equation}
\label{eq:block_risk_transform}
V_{Q_k}(\theta)
=
1-(1-V_P(\theta))^k.
\end{equation}

If we further define $\Phi_k(v)=1-(1-v)^k$, we see that $\Phi_k$ is strictly increasing
on $[0,1]$ and $V_P(\theta)>\epsilon
\iff
V_{Q_k}(\theta)>\Phi_k(\epsilon)$. 
\end{prop}

\noindent For the sample-feasible decision
family $\{A_m\}_{m\ge0}$, we recall
the support-subsample certificate
of \citet{Campi2018GenScen} in order to show how it behaves under our block transformation. For an index set
$I=\{i_1<\cdots<i_r\}
\subseteq\{1,\ldots,N\}$, denote an ordered primitive sublist as $\omega_{N,I}
:=
(\delta_{i_1},\ldots,\delta_{i_r})$. As defined in \citet{Campi2018GenScen}, a support subsample is an ordered sublist
indexed by $I\subseteq\{1,\ldots,N\}$ such that $A_{|I|}(\omega_{N,I})=A_N(\omega_N)$. Let $B_N$ be a measurable selector of such
a sublist, with random cardinality
$r_N=|B_N(\omega_N)|$, the the equal-allocation support-subsample
threshold of \citet{Campi2018GenScen} is then given by 
\begin{equation}
\label{eq:equal_alloc}
\varepsilon_N^{\mathrm{SS}}(r_N)
=
1-
\left[
\frac{\beta}{N\binom {N}{r_N}}
\right]^{1/(N-r_N)},
\quad 0\le r_N<N,
\end{equation}
with $\varepsilon_N^{\mathrm{SS}}(N)=1$. This threshold satisfies the product measure violation bound 
\begin{equation}
P^{\otimes N}
\left\{
V_P(A_N(\omega_N))>
\varepsilon_N^{\mathrm{SS}}(r_N)
\right\}
\le\beta.
\end{equation}

In order to compare the behaviour of this certificate under grouping, we require the following two properties to hold under our exact-block reprsentation. For each exact block size $k$,
let $
A_m^{(k)}:
(\Delta^k)^m\to\Theta$ 
be a measurable grouped decision
family. Further, assume that $A_m^{(k)}$ is
sample-feasible for every
$m\ge1$ and admits a measurable
reproducing-subsample selector
$B_m^{(k)}$. For the full grouped bank,
define $I_k
:=
B_{M_k}^{(k)}(\Omega_k), \; r_k=|I_k|$. Letting $\Omega_{k,I_k}$
be the ordered sublist
selected by $I_k$, we require
\begin{equation}
\label{eq:grouped_reproduction}
A_{r_k}^{(k)}
(\Omega_{k,I_k})
=
A_{M_k}^{(k)}(\Omega_k)
=
A_N(\omega_N).
\end{equation} 

The first equality in \eqref{eq:grouped_reproduction} establishes support-subsample reproduction, while the second establishes decision equivalence between the primitive and grouped representations. Under the i.i.d. block law \(Q_k=P^{\otimes k}\), we may therefore apply the support-subsample certificate to the grouped decision family while retaining the same primitive decision. By Proposition~\ref{prop:block_risk_transform}, the primitive violation tolerance \(\epsilon\) corresponds exactly to the grouped tolerance \(\Phi_k(\epsilon)=1-(1-\epsilon)^k\). Consequently, for \(0\le r<M_k\), the grouped certificate is sufficient to certify the primitive-risk target whenever
\begin{equation}
\begin{gathered}
\varepsilon_{M_k}^{\mathrm{SS}}(r)
\le \Phi_k(\epsilon)
\\[1pt]
\Updownarrow
\\[1pt]
1-
\left(
\frac{\beta}
{M_k\binom{M_k}{r}}
\right)^{\frac{1}{M_k-r}}
\le 1-(1-\epsilon)^k
\\[1pt]
\Updownarrow
\\[1pt]
M_k\binom{M_k}{r}
(1-\epsilon)^{k(M_k-r)}
\le \beta.
\end{gathered}
\end{equation}
Substituting \(M_k=N/k\), we define the \emph{elementary certificate factor}
\begin{equation}
\Psi_k(N,r)
:=
\frac Nk
\binom{N/k}{r}
(1-\epsilon)^{N-kr}.
\end{equation} For a realized grouped sample bank, the certificate is evaluated at the random reproducing cardinality \(r_k\). Thus, \(\Psi_k(N,r_k)\le\beta\), with \(r_k<M_k\), is equivalent to requiring that the grouped support-subsample threshold does not exceed the grouped risk tolerance corresponding to \(\epsilon\).

Using the above definition of a certificate factor, we provide the following proposition which shows that the certificate factor controls the probability of incorrectly certifying a decision whose primitive violation probability exceeds \(\epsilon\).

\begin{prop}[Primitive-risk certification
under exact grouping]
\label{prop:prim_risk_cert_exact_grouping}
Under the preceding setup, define $\mathrm{CERT}_k
=
\{r_k<M_k,\,
\Psi_k(N,r_k)\le\beta\}$. Then
\begin{equation}
\label{eq:selective_primitive_certificate}
P^{\otimes N}
\left\{
\mathrm{CERT}_k
\land
V_P(A_N(\omega_N))>\epsilon
\right\}
\le\beta.
\end{equation}
\end{prop}

\subsection{Representation dependence}
\label{sec:representation_dependence}

Having established the tools with which to describe certificate behaviour, we now show that the elementary certificate is representation dependent. We do so using the basic example of a scalar scenario program which seeks the smallest scalar larger than all samples. Let \(F\) be the continuous CDF of \(P\), and consider the scalar scenario program
\[
\min_{\theta\in\mathbb R}\ \theta
\qquad
\text{subject to}\qquad
f(\theta,\delta_i)=\delta_i-\theta\le0,
\quad i=1,\ldots,N.
\] The optimizer and primitive feasible sets have the form $\theta_N=\max_i\delta_i,$ and $
\Theta_\delta=[\delta,\infty)$ respectively. For the $k$-block representation, we define block representation 
\begin{equation}
Y_j^{(k)}
:=
\max_{1\le r\le k}
\delta_{(j-1)k+r}.
\end{equation}

Clearly, the grouped program minimizes $\theta$
subject to
$\theta\ge Y_j^{(k)}$ for all $j$.
Its optimizer is $\theta_{M_k}^{(k)}
=
\max_jY_j^{(k)}
=
\theta_N$. We now present our first theorem concerning representation dependence for the scalar scenario-program.

\begin{thm}[Quantitative representation dependence]
\label{thm:quant_rep_dep_for_scal_max}
For each representation, choose the
measurable support selector that retains
the least-index block attaining
the full-bank maximum. For the scalar program, every proper
exact block size $k\mid N$, $k<N$,
satisfies:

\emph{(i) Decision invariance:}
$\theta_{M_k}^{(k)}=\theta_N$.

\emph{(ii) Exact primitive-risk law:} $P^{\otimes N}
\{V_P(\theta_N)>\epsilon\}
=
(1-\epsilon)^N$.

\emph{(iii) Reproducing cardinality:}
$r_k=1$ almost surely.

\emph{(iv) Certificate-factor distortion:}
\begin{equation}
\label{eq:certificate_distortion}
R_\epsilon(k)
:=
\frac{\Psi_1(N,1)}
     {\Psi_k(N,1)}
=
k^2(1-\epsilon)^{k-1}.
\end{equation}
\end{thm}

\noindent Theorem~\ref{thm:quant_rep_dep_for_scal_max}
quantifies how the elementary certificate
changes under exact grouping. Two consequences follow. First, the discrepancy can become
arbitrarily large as the prescribed
violation tolerance decreases.
Maximizing the distortion over
unrestricted integer block sizes gives $\max_{k\in\mathbb N}R_\epsilon(k)
\sim
4e^{-2}\epsilon^{-2},
\; \epsilon\downarrow0$.
A complete derivation appears
in Appendix~\ref{app:proofs}. Second, this discrepancy can produce
opposite certificate-acceptance decisions
for the same primitive observations
and optimizer.
From Theorem~\ref{thm:quant_rep_dep_for_scal_max}, we have the following corollary concretely showing the change in certification behaviour. Unlike the preceding fixed-budget
comparison, the following construction
considers a family of predeclared designs
indexed by the risk tolerance $\epsilon$. Accordingly, $N_\epsilon$ denotes the
primitive sample budget selected for
that tolerance.

\begin{cor}[Fixed-confidence certification reversal]
\label{cor:fix_conf_cert_rev}

Fix $\beta\in(0,1)$ and choose
an integer $M\ge2$ satisfying $M^2e^{-2(M-1)}<\beta$. For $\epsilon\in(0,1)$,
choose $k_\epsilon\in
\operatorname*{arg\,max}_{k\in\mathbb N}
R_\epsilon(k), \;
N_\epsilon=Mk_\epsilon$. These choices are made
before observing the sample bank. For all sufficiently small
$\epsilon>0$,
\begin{equation}
\Psi_{k_\epsilon}(N_\epsilon,1)
<
\beta
<
\Psi_1(N_\epsilon,1).
\end{equation}
\end{cor}

\noindent For illustration, let
$\epsilon=0.05$, $\beta=0.01$,
$k=39$, and $N=195$.
Then $\Psi_{39}(195,1)\approx0.00837
<
\beta
<
1.81321
\approx\Psi_1(195,1)$. Thus despite the losslessly equivalent representations of the primitives, the certificate under grouping accepts while
the singleton-fed certificate rejects.

\subsection{Certificate implication under coarsening}
\label{sec:coarsening}

Having established that losslessly equivalent scenario representations can yield different certificate assessments, we now investigate conditions under which these differences admit a predictable ordering. Specifically, we derive sufficient conditions under which certification with a finer scenario representation implies certification under a coarser representation of the same primitive observations.

Consider two nested exact representations with fine block size \(k_a\) and coarse block size \(k_b=mk_a\), where \(m\ge2\) is an integer and both block sizes divide \(N\). Let \(M_a=N/k_a\) and \(M_b=N/k_b\), so that \(M_a=mM_b\). Each coarse block therefore contains exactly \(m\) consecutive fine blocks. Using one-based block indices,
define the parent-block mapping $\pi(j)
=
1+\left\lfloor
\frac{j-1}{m}
\right\rfloor,
\;
j\in\{1,\ldots,M_a\}$.

For reproducing sublists $C_a$ and $C_b$
of the fine and coarse representations,
respectively, write
$r_a:=|C_a|$ and $r_b:=|C_b|$.
We call these sublists
\emph{coarsening compatible} if $C_b\subseteq\pi(C_a)$. This condition implies $r_b\le r_a$.
\begin{thm}[Elementary-certificate dominance
under nested coarsening]
\label{thm:elem_cert_dom}

Fix $\epsilon\in(0,1)$ and write
$q=1-\epsilon$. Let $h\in\mathbb N$ be a deterministic,
sample-independent upper bound on
the reproducing-sublist cardinality. Then suppose the fine and coarse sublists
are coarsening compatible and satisfy $r_b\le r_a\le h<M_b$.  
If $ mq^{k_b-k_a}\ge1,
\; h-1\le
\frac{M_b-q^{k_b}}{1+q^{k_b}}$,
then
\begin{equation}
\Psi_{k_b}(N,r_b)
\le
\Psi_{k_b}(N,r_a)
<
\Psi_{k_a}(N,r_a).
\end{equation} Consequently, $\mathrm{CERT}_{k_a}
\implies
\mathrm{CERT}_{k_b}$. Where $C_a$ and $C_b$ are used to evaluate the corresponding certificate predicates.
\end{thm}

\subsection{Primitive-envelope certification without block accounting}

The preceding results demonstrate that application of the equal-allocation
support-subsample certificate can produce
different assessments under losslessly
equivalent atomic scenario representations. However, as lossless grouping admits exact
decoding, any fixed primitive-bank
certificate can be made invariant to
grouping by evaluating it on the
recovered primitive observations. The more consequential question is therefore
whether certification on the primitive
bank can provide sufficiently informative
risk assessments without relying on
grouping. We investigate this question using the
nondegeneracy-free Wait-and-Judge
certificate of
\citet{Garatti2025NonconScenOpt},
applied to coordinatewise scenario
envelopes.

In order to use the WJ certificate, we introduce coordinatewise scenario envelopes. This construction serves two purposes: it preserves the sampled constraint envelope under exact regrouping, and it defines a sample-feasible decision family with a bounded reproducing support subsample. Moreover, the envelope decision family satisfies the consistency conditions of \citet{Garatti2025NonconScenOpt}.

\begin{defin}[Coordinatewise scenario envelope]
\label{def:coord_scen_env}

Let \(G\in\mathbb N\) denote the number of scalar scenario-score functions, and let \(H:\Delta\to\overline{\mathbb R}^{G}\) be a measurable vector-valued map, with components \(H_g:\Delta\to\overline{\mathbb R}\) for \(g=1,\ldots,G\) where
$\overline{\mathbb R}
=\mathbb R\cup\{-\infty,+\infty\}$. Each component \(H_g(\delta)\), referred to as a score coordinate, represents a scalar quantity associated with primitive observation \(\delta\). Write $a\preceq b$ for coordinatewise
inequality and $a\vee b$ for
coordinatewise maximum. Further, define the bottom envelope
$\bot_G:=(-\infty,\ldots,-\infty)$ and for $N\ge1$, let $E_N=\bigvee_{i=1}^{N}H(\delta_i),
\; E_0=\bot_G$. The envelope coverage set is $\Gamma(E) = \{\delta:H(\delta)\preceq E\}$, and the
exceedance probability is
\begin{equation}W_P(E) =
P\{\delta:H(\delta)\npreceq E\}.
\label{eq:envelope_exceedance}
\end{equation}
\end{defin}

\noindent To connect the preceding representation
results with a structure-aware
certificate, we establish two properties
of coordinatewise envelopes.  

\begin{prop}[Envelope invariance and canonical witnesses]
\label{prop:env_part_invar}

Let $\Pi$ be an exact partition
of $\{1,\ldots,N\}$ and define $E_{N,\Pi}
:=
\bigvee_{B\in\Pi}
\bigvee_{i\in B}H(\delta_i)$. Then $E_{N,\Pi}=E_N$. Let $\mathcal A_N
:=
\{g\in\{1,\ldots,G\}:
E_{N,g}>-\infty\}$ be the active coordinate set. For each $g\in\mathcal A_N$,
let $i_g^\star$ be the least
primitive index attaining
$E_{N,g}$. Define $C_{\mathrm{prim}}
=
\{i_g^\star:
g\in\mathcal A_N\},
\;
r_{\mathrm{prim}}
=
|C_{\mathrm{prim}}|$. The selected primitive observations
reproduce the envelope $\bigvee_{i\in C_{\mathrm{prim}}}
H(\delta_i)
=
E_N,
\;
r_{\mathrm{prim}}\le G$. For any exact block size $k$,
define
\[
C_k
=
\left\{
1+\left\lfloor
\frac{i_g^\star-1}{k}
\right\rfloor:
g\in\mathcal A_N
\right\}.
\] Then $C_k$ is a reproducing
sublist of block indices. For nested block sizes
$k_b=mk_a$, $C_{k_b}=\pi(C_{k_a}),
\;
|C_{k_b}|
\le|C_{k_a}|\le G$.
\end{prop}

\noindent Having defined the scenario envelope and shown it is invariant under exact regrouping and can be reproduced by at most (G) primitive observations, we introduce and apply the WJ certificate. For a decision family
$\{\mathbf M_m\}_{m\ge0}$,
the so-called ``consistency property" or \citet{Garatti2025NonconScenOpt} requires permutation
invariance, confirmation under
appropriate additions, and
responsiveness to inappropriate
additions. The coordinatewise envelope decision family satisfies these requirements because its maximum is invariant to observation ordering, unchanged by additional scores bounded by the existing envelope, and modified whenever a new score exceeds it in at least one coordinate.

Under the consistency property, Theorem~4 of \citet{Garatti2025NonconScenOpt} provides a distribution-free violation threshold. For each deterministic integer
$r\in\{0,\ldots,N-1\}$, define
$\varepsilon_N^{\mathrm{WJ}}(r)=1-t_N(r)$, 
where $t_N(r)\in(0,1)$ solves
\begin{equation}
\label{eq:wj_root}
\frac{\beta}{N}
\sum_{m=r}^{N-1}\binom mr t^{m-r}
-
\binom Nr t^{N-r}=0,
\end{equation}
and set $\varepsilon_N^{\mathrm{WJ}}(N)=1$. The theorem then guarantees that the probability of the returned decision's violation probability exceeding the support-dependent threshold is at most \(\beta\). In particular, when applied to the primitive envelope decision family defined below, for which \(\mathbf M_N(\omega_N)=E_N\) and the decision violation probability is \(W_P(E_N)\), the theorem yields
\begin{equation}
\label{eq:wj_original_guarantee}
P^{\otimes N}
\left\{
W_P(E_N)>
\varepsilon_N^{\mathrm{WJ}}(s_N^\star)
\right\}
\le\beta.
\end{equation}
This guarantee uses the minimum irreducible support-list cardinality \(s_N^\star\). The following proposition establishes the required consistency property for the envelope decision family and shows that the bound remains valid when \(s_N^\star\) is replaced by the explicitly constructed canonical witness cardinality \(r_{\mathrm{prim}}\). By certifying the envelope using the original primitive observations and their canonical reproducing witness, we obtain a certification procedure whose assessment is unchanged by subsequent exact grouping.
\begin{prop}[Primitive WJ envelope certification]
\label{prop:prim_wj_env_cert}
Let
$\mathcal Z=\overline{\mathbb R}^{G}$
be the envelope decision space
and define $\mathcal Z_\delta
=
\{E\in\mathcal Z:
H(\delta)\preceq E\}$. For every $m\ge0$, define the
measurable envelope decision map $\mathbf M_m:\Delta^m\to\mathcal Z$,
where $\mathbf M_m(\delta_{1:m})
=
\bigvee_{i=1}^{m}H(\delta_i),
\;
\mathbf M_0(\varnothing)=\bot_G$. This family satisfies the consistency property of
\citet{Garatti2025NonconScenOpt}, and its minimum irreducible
support-list cardinality
$s_N^\star$ satisfies $s_N^\star
\le r_{\mathrm{prim}}\le G$. 
Since the WJ threshold is nondecreasing
in its support-cardinality argument, $
\varepsilon_N^{\mathrm{WJ}}(s_N^\star)
\le
\varepsilon_N^{\mathrm{WJ}}(r_{\mathrm{prim}})$. Consequently,
\begin{equation}
\label{eq:primitive_wj_certificate}
P^{\otimes N}
\left\{
W_P(E_N)>
\varepsilon_N^{\mathrm{WJ}}
(r_{\mathrm{prim}})
\right\}
\le\beta.
\end{equation}
\end{prop}

\noindent This result demonstrates that representation-dependent certification is not unavoidable. By evaluating the WJ certificate directly on the primitive envelope and its canonical witness, the certification assessment remains unchanged under subsequent exact regrouping of the same observations. Unlike the elementary support-subsample certificate, this procedure does not depend on the selected block size. This is a property of the specified
primitive-WJ procedure, not a claim that WJ
evaluated on grouped atomic observations is
intrinsically representation invariant.

\section{Application to Scenario MPC}
\label{sec:mpc}

To illustrate the practical implications of
representation-dependent certification, we consider
scenario-based MPC for
mobile robot collision avoidance. Our application
uses fixed-normal collision constraints, whose
sampled right-hand sides admit the coordinatewise
envelope construction developed in
Section~\ref{sec:derivation}. Throughout, we fix context $c$ before sampling and represent the collection of collision constraint coordinates as $\mathcal{G}_c$. Further, we use $a_{c,g}$ to denote fixed
projection normals and $t(g)$ prediction
stages. Under these conventions, a primitive observation $\delta$
imposes the modeled constraints $a_{c,g}^{\top}p_{t(g)}(\theta)
\ge b_{c,g}(\delta),
\; g\in\mathcal G_c$,
where $p_{t(g)}(\theta)$ denotes the predicted
ego position under decision $\theta$.

We construct measurable score coordinates
$H_{c,g}(\delta)$ that conservatively
upper-bound the corresponding offsets
$b_{c,g}(\delta)$ on the supported domain. The sampled
envelope is $E_N^{(c)}
=
\bigvee_{i=1}^{N}H_c(\delta_i)$. A common, certificate-independent numerical
feasibility gate requires the MPC proposal
$\theta_N$ to satisfy all deterministic
admissibility checks and the verified
envelope constraints $a_{c,g}^{\top}p_{t(g)}(\theta_N)
\ge E_{N,g}^{(c)}$ for every group with a finite envelope
coordinate. On this event, a fresh primitive
observation covered by the envelope cannot
violate any modeled collision constraint.
Consequently,
\begin{equation}
\label{eq:main_mpc_transfer}
V_{P_c}(\theta_N)
\le W_{P_c}(E_N^{(c)}).
\end{equation}
Thus, when the MPC controller satisfies the numerical feasibility checks, the statistical guarantee established for the sampled scenario envelope also provides a guarantee on the controller's modeled collision-constraint violation probability. 

We compare certificate acceptance on identical
primitive scenario banks and MPC proposals,
holding the sampled envelope and numerical
feasibility checks fixed across methods.
The experiments measure certification
availability rather than changes in
primitive violation probability or
closed-loop collision safety.
Additional mathematical and experimental details are provided in
Appendix~\ref{app:experiments}.

\subsection{Results}

\begin{figure}[t]
\centering
\includegraphics[width=\linewidth]
{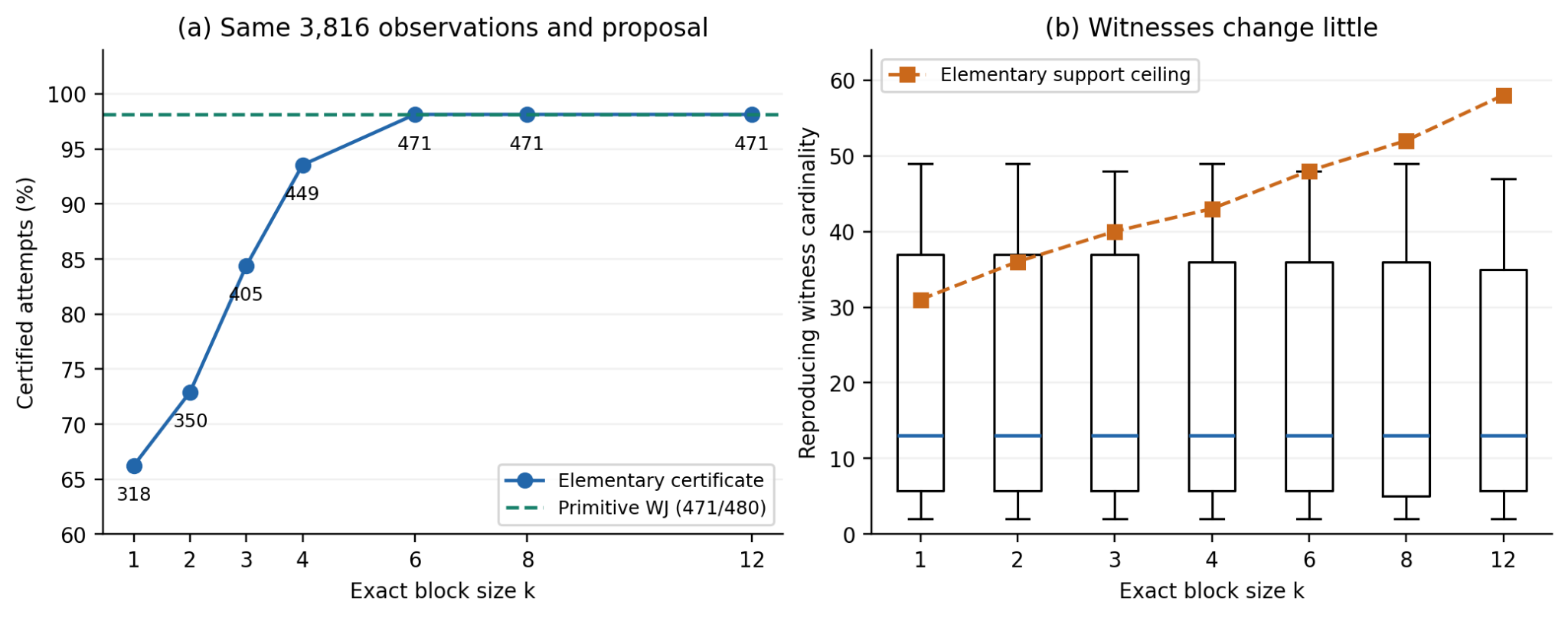}
\caption{
Certification under exact regrouping of 480 paired
MPC attempts ($N=3816$, $\epsilon=0.05$,
$\beta=0.01$).
(a) Elementary-certificate availability varies
with block size; primitive WJ is unchanged.
(b) Reproducing-witness sizes and elementary
support ceilings.
All common nonrisk rejections remain included.
}
\label{fig:same_bank_certification}
\end{figure}

\begin{table}[t]
\centering
\begin{tabular}{lccc}
\hline
Certificate
& Primary $N=3816$
& Fresh $N=1536$
& First $\ge95\%$\\
\hline
SS, $k=1$
& 318/480
& 218/480
& 5760\\
SS, $k=12$
& 471/480
& 318/480
& 3072\\
Primitive WJ
& 471/480
& 476/480
& 1536\\
\hline
\end{tabular}
\caption{
Certification availability for the
primary paired-bank experiment and
a separate varying scenario-budget experiment.
The final column reports
the first tested primitive budget
attaining at least 95\% certification.
}
\label{tab:certificate_comparison}
\end{table}

The primary experiment holds the primitive observations, envelopes, MPC proposals, and nonrisk gates fixed across certificates. At \(N=3816\), singleton elementary certification accepts 318/480 attempts, compared with 471/480 under \(k=12\) grouping, an increase of 31.9 percentage points (Fig.~\ref{fig:same_bank_certification}). The paired outcomes comprise 318 accepted by both methods, 153 accepted only under grouping, none accepted only under singleton accounting, and nine rejected by both. Importantly, \(r_{12}\le r_{\mathrm{prim}}\le49\), below the grouped certificate's acceptance ceiling of 58. Thus, all 471 gate-passing attempts satisfy grouped certification without requiring any reduction in witness cardinality. The improvement arises from certificate accounting rather than changes in sampled information or MPC optimization.

Primitive WJ certification accepts the same 471 primary attempts without grouping. On separate fresh banks at \(N=1536\), singleton elementary, grouped elementary, and primitive WJ certification accept 218, 318, and 476 of 480 attempts, respectively. Their first tested budgets attaining at least 95\% certification availability are 5760, 3072, and 1536 (Table~\ref{tab:certificate_comparison}). These results concern certification availability, not minimum sample requirements or measured collision-risk improvements.

\section{Conclusion}

We demonstrated that losslessly equivalent scenario representations can yield different acceptance decisions under an elementary support-subsample certificate, despite preserving the sampled information, optimizer, and primitive risk. We quantified this discrepancy for a scalar scenario program and established sufficient conditions for certificate dominance under nested coarsening. We further showed that primitive Wait-and-Judge certification of coordinatewise envelopes can provide stronger assessments without grouped-scenario accounting. In paired scenario-MPC experiments, grouping yielded 153 additional certifications without requiring witness reduction, while primitive WJ accepted the same attempts. These findings highlight the importance of certificate accounting when evaluating scenario sample efficiency. Future lines of work include investigating other forms of certificates for their behaviour under grouping, analysing grouping behaviour for more complex scenario programs, and research into the practical implications of sample grouping for grouping sensitive certificates. 

\acks{Throughout this publication, basic proof formulation was assisted by ChatGPT GPT-6. Code implementation, execution, and figure generation used ChatGPT GPT-6-astra with manual review by the authors. All math was verified manually and reformatted by the authors. 

This material is based upon work supported by the National Science Foundation CISE Graduate Fellowships under Grant No. 2313998. Any opinions, findings, and conclusions or recommendations expressed in this material are those of the author(s) and do not necessarily reflect the views of the National Science Foundation.}

\bibliography{main}

\appendix
\section{Proofs of the theoretical results}
\label{app:proofs}

\subsection{Scenario decisions and primitive risk}

\begin{proof}[Proof of feasible-region identity~\eqref{eq:feasible_region_identity}]
For the predeclared exact partition, the list of indices
$(j-1)k+r$, with $j=1,\ldots,M_k$ and $r=1,\ldots,k$,
contains every element of $\{1,\ldots,N\}$ exactly once. Hence
\[
\bigcap_{j=1}^{M_k}\Theta_{Z_j^{(k)}}
=\bigcap_{j=1}^{M_k}\bigcap_{r=1}^{k}
\Theta_{\delta_{(j-1)k+r}}
=\bigcap_{i=1}^{N}\Theta_{\delta_i}.
\]
\end{proof}

\begin{proof}[Proposition~\ref{prop:block_risk_transform}]
For a fixed decision $\theta$, the event that its block is satisfied
is the intersection of the $k$ independent primitive satisfaction
events. Consequently,
\[
Q_k\{Z:\theta\in\Theta_Z\}
=\prod_{r=1}^{k}P\{\delta:\theta\in\Theta_\delta\}
=(1-V_P(\theta))^k.
\]
Taking complements gives
$V_{Q_k}(\theta)=1-(1-V_P(\theta))^k=\Phi_k(V_P(\theta))$.
\end{proof}

\begin{proof}[Proposition~
\ref{prop:prim_risk_cert_exact_grouping}]
Write $q=1-\epsilon$ and
$\theta_N:=A_N(\omega_N)
=A_{M_k}^{(k)}(\Omega_k)$.

\noindent For $r_k<M_k$, the elementary
acceptance condition is equivalent to
\[
\begin{aligned}
\Psi_k(N,r_k)\le\beta
\iff\
\varepsilon_{M_k}^{\mathrm{SS}}(r_k)
\le\Phi_k(\epsilon).
\end{aligned}
\]

\noindent By the strictly increasing
block-risk transformation,
\[
\mathrm{CERT}_k
\cap\{V_P(\theta_N)>\epsilon\} \subseteq
\left\{
V_{Q_k}(\theta_N)>
\varepsilon_{M_k}^{\mathrm{SS}}(r_k)
\right\}.
\]

\noindent The exact representation $R_k$
transports $P^{\otimes N}$
to $Q_k^{\otimes M_k}$. By the assumed sampled feasibility,
measurability, and reproducing-sublist
properties of $A_m^{(k)}$ and
$B_m^{(k)}$, the
support-subsample theorem of \citet{Campi2018GenScen} bounds the
probability of the latter event by
$\beta$.
\end{proof}

\subsection{Representation dependence}

\begin{proof}[Theorem~\ref{thm:quant_rep_dep_for_scal_max}]
Because the optimizer of the scalar program is the primitive maximum,
while the grouped optimizer is the maximum of all block maxima, the optimizers agree for every primitive bank.
Let $F$ be the continuous CDF of $P$. By the probability integral
transform, $U_i=F(\delta_i)$ are i.i.d. drawn from 
$\operatorname{Uniform}(0,1)$. Since
$V_P(\theta_N)=1-F(\theta_N)$, for $q=1-\epsilon$,
\[
\begin{aligned}
P^{\otimes N}\{V_P(\theta_N)>\epsilon\}
&=P^{\otimes N}\{\max_{i}U_i<q\}\\
&=q^N.
\end{aligned}
\]
Continuity makes the maximizing primitive index unique almost surely.
Keeping only its primitive observation or its containing exact block
reproduces the full solution, giving $r_k=1$ almost surely.
Finally, $\Psi_k(N,1)=(N/k)^2q^{N-k}$ and
$\Psi_1(N,1)=N^2q^{N-1}$; their ratio is
$k^2q^{k-1}$.
\end{proof}

\begin{proof}[Derivation of the untruncated
distortion asymptotic]
Let $q=1-\epsilon$ and
$\lambda=-\log q$.
Consider the continuous extension $R_\epsilon(x)=x^2q^{x-1},
\; x\ge1$. Its logarithm satisfies
\[
\frac{d}{dx}\log R_\epsilon(x)
=\frac{2}{x}-\lambda,
\qquad
\frac{d^2}{dx^2}\log R_\epsilon(x)
=-\frac{2}{x^2}<0.
\]
For sufficiently small $\epsilon$,
the unique continuous maximizer is
$x_\epsilon^\star=2/\lambda$.
Its value is
\[
R_\epsilon(x_\epsilon^\star)
=
\frac{4e^{-2}}{q\lambda^2}.
\]

\noindent By unimodality, an integer maximizer
$k_\epsilon$ belongs to
$\{\lfloor x_\epsilon^\star\rfloor,
\lceil x_\epsilon^\star\rceil\}$.
Consequently, $\lambda k_\epsilon\to2,
\;
\frac{R_\epsilon(k_\epsilon)}
{R_\epsilon(x_\epsilon^\star)}
\to1$ as $\epsilon\downarrow0$.
Since $\lambda\sim\epsilon$
and $q\to1$, we conclude that
\[
\max_{k\in\mathbb N}R_\epsilon(k)
\sim4e^{-2}\epsilon^{-2}.
\]
\end{proof}

\begin{proof}[Corollary~\ref{cor:fix_conf_cert_rev}]
Fix $\beta\in(0,1)$. Choose a fixed integer $M\ge2$
sufficiently large that $M^2e^{-2(M-1)}<\beta$. Such an integer exists because
$M^2e^{-2(M-1)}\to0$
as $M\to\infty$. Let $\lambda_\epsilon
=
-\log(1-\epsilon)$
and select
\[
k_\epsilon\in
\operatorname*{arg\,max}_{k\in\mathbb N}
k^2(1-\epsilon)^{k-1}.
\]

\noindent Set
$N_\epsilon=Mk_\epsilon$.
These choices are made using
$\epsilon$ and $\beta$, before
observing any primitive scenarios. By the preceding asymptotic
maximization argument, $k_\epsilon\to\infty,
\;
\lambda_\epsilon k_\epsilon\to2
\quad\text{as }\epsilon\downarrow0$. For grouped accounting,
\[
\begin{aligned}
\Psi_{k_\epsilon}(N_\epsilon,1)
&=
M^2(1-\epsilon)^{
k_\epsilon(M-1)}\\
&=
M^2
e^{-\lambda_\epsilon
k_\epsilon(M-1)}\\
&\longrightarrow
M^2e^{-2(M-1)}<\beta.
\end{aligned}
\]

\noindent For singleton accounting,
\[
\begin{aligned}
\Psi_1(N_\epsilon,1)
&=
M^2k_\epsilon^2
(1-\epsilon)^{Mk_\epsilon-1}\\
&=
M^2k_\epsilon^2
e^{-\lambda_\epsilon
(Mk_\epsilon-1)}\\
&\longrightarrow\infty.
\end{aligned}
\]

\noindent Both strict inequalities therefore
hold for all sufficiently small
$\epsilon>0$.
\end{proof}

\subsection{Certificate implication under coarsening}

\begin{proof}[Theorem~\ref{thm:elem_cert_dom}]
Write $q=1-\epsilon$,
$M_a=N/k_a=mM_b$. For $0\le r<M_b-1$, the ratio of
successive coarse factors is
\[
\frac{\Psi_{k_b}(N,r+1)}{\Psi_{k_b}(N,r)}
=\frac{M_b-r}{r+1}q^{-k_b}.
\]
It is at least one exactly when
$r\le(M_b-q^{k_b})/(1+q^{k_b})$.
The hypothesis on $h-1$, together with
$r_b\le r_a\le h<M_b$, gives
$\Psi_{k_b}(N,r_b)\le\Psi_{k_b}(N,r_a)$.
For $0\le r\le h$, using $M_a=mM_b$,
\[
\begin{aligned}
\frac{\Psi_{k_a}(N,r)}{\Psi_{k_b}(N,r)}
&=m\frac{\binom{mM_b}{r}}{\binom{M_b}{r}}
  q^{(k_b-k_a)r}\\
&=m\prod_{i=0}^{r-1}\frac{mM_b-i}{M_b-i}
  q^{(k_b-k_a)r}\\
&\ge m(mq^{k_b-k_a})^r\ge m>1.
\end{aligned}
\]
Apply this bound at $r=r_a$ and combine with coarse-factor
monotonicity to obtain the theorem.
\end{proof}

\subsection{Primitive-envelope certification without block accounting}

\begin{proof}[Proposition~\ref{prop:env_part_invar}]
Coordinatewise maximum is associative
and commutative. Hence, for every
exact partition $\Pi$,
\[
E_{N,\Pi}
=
\bigvee_{B\in\Pi}
\bigvee_{i\in B}H(\delta_i)
=
E_N.
\]

\noindent The coverage sets and exceedance
probabilities are therefore identical. For each active coordinate
$g\in\mathcal A_N$, the index
$i_g^\star$ attains the sampled
maximum $E_{N,g}$. Because $i_g^\star\in C_{\mathrm{prim}}$,
\[
\bigvee_{i\in C_{\mathrm{prim}}}
H(\delta_i)
=
E_N.
\]

\noindent The equality also holds when
$\mathcal A_N=\varnothing$, using
the empty-maximum convention
$\bigvee\varnothing=\bot_G$. At most one index is selected per
active coordinate, so
$r_{\mathrm{prim}}\le G$. For block size $k$, define
\[
\operatorname{blk}_k(i)
=
1+\left\lfloor\frac{i-1}{k}\right\rfloor.
\]

\noindent Every selected block
$\operatorname{blk}_k(i_g^\star)$
contains a primitive observation
attaining coordinate maximum $E_{N,g}$.
Consequently, $C_k$ reproduces $E_N$. For nested sizes $k_b=mk_a$,
\[
\begin{aligned}
\operatorname{blk}_{k_b}(i)
&=
1+\left\lfloor
\frac{i-1}{mk_a}
\right\rfloor\\
&=
1+\left\lfloor
\frac{
\left\lfloor(i-1)/k_a\right\rfloor
}{m}
\right\rfloor\\
&=
\pi\bigl(
\operatorname{blk}_{k_a}(i)
\bigr).
\end{aligned}
\]

\noindent Applying this identity to all
canonical maximizing indices gives
\[
C_{k_b}=\pi(C_{k_a}),
\qquad
|C_{k_b}|\le|C_{k_a}|\le G.
\]
\end{proof}

\begin{proof}[Proposition~\ref{prop:prim_wj_env_cert}]
Set $\mathcal Z=\overline{\mathbb R}^{G}$ and
$\mathcal Z_\delta=\{E:H(\delta)\preceq E\}$.
For $m\ge0$, the decision map
$\mathbf M_m(\delta_{1:m})=\bigvee_{i=1}^{m}H(\delta_i)$
is invariant under permutations.
If every newly added score is dominated coordinatewise by
$E_m=\mathbf M_m(\delta_{1:m})$, then their maximum with $E_m$
remains $E_m$.
If at least one added score is not dominated by $E_m$,
some coordinate strictly increases, so the decision changes.
These are Property 1(i)--(iii) of
\citet{Garatti2025NonconScenOpt}.
Moreover,
\[
V_{\mathrm{WJ}}(E)
=P\{\delta:E\notin\mathcal Z_\delta\}
=P\{\delta:H(\delta)\npreceq E\}=W_P(E).
\]
The canonical primitive witness reproduces $E_N$.
Starting from this finite witness, remove an
index whenever its removal preserves the
coordinatewise envelope, and repeat until
no further removal is possible. The resulting
sublist is irreducible and reproduces $E_N$.
Since $s_N^\star$ is the minimum cardinality
among all irreducible reproducing sublists, $s_N^\star\le r_{\mathrm{prim}}$. 

\noindent For completeness, the WJ threshold is
nondecreasing in cardinality. For $0\le r<N$, define
\[
T_r(t)
:=
\frac{\beta}{N}
\sum_{m=r}^{N-1}
\frac{\binom mr}{\binom Nr}
t^{m-N},
\qquad 0<t\le1.
\]Equation~\eqref{eq:wj_root} is equivalent to $T_r(t_N(r))=1$. Every $T_r$ is strictly decreasing in $t$. For $r<N-1$,
\[
\begin{aligned}
T_{r+1}(t)
&=
\frac{\beta}{N}
\sum_{m=r+1}^{N-1}
\frac{m-r}{N-r}
\frac{\binom mr}{\binom Nr}
t^{m-N}
\\
&<T_r(t).
\end{aligned}
\] It follows that $t_N(r+1)<t_N(r)$ and therefore
$\varepsilon_N^{\mathrm{WJ}}(r+1)>
\varepsilon_N^{\mathrm{WJ}}(r) $for $r<N-1$. The convention
$\varepsilon_N^{\mathrm{WJ}}(N)=1$
preserves nondecreasingness. Consequently, $\varepsilon_N^{\mathrm{WJ}}(s_N^\star)
\le
\varepsilon_N^{\mathrm{WJ}}(r_{\mathrm{prim}})$. The event in
\eqref{eq:primitive_wj_certificate}
is contained in
\[
\left\{
W_P(E_N)>
\varepsilon_N^{\mathrm{WJ}}(s_N^\star)
\right\},
\]
which has probability at most $\beta$
by Theorem~4 of
\citet{Garatti2025NonconScenOpt}.

\end{proof}

\section{Experimental implementation and reproducibility}
\label{app:experiments}

This appendix describes the
scenario-MPC benchmark, primitive
sampling law, fixed-normal constraint
construction, numerical feasibility
verification, and certificate
evaluation procedure. The experimental design isolates
certificate accounting by holding
each primitive bank, sampled envelope,
numerical proposal, and nonrisk gate
fixed across certificate arms. All encounter contexts and
constraint normals are determined
before evaluation sampling.

\subsection{Experimental design}
\label{app:experiment_question}

We model a typical scenario-based MPC problem in the form of mobile robot navigation. Six encounter families combine straight
and curved roads with three obstacle configurations: one obstacle with one trajectory model, one obstacle with three trajectory models, and three obstacles with three trajectory models. This provides a variety of interaction settings and can be considered a reasonably interesting application. For each encounter family, we define a subset of four encounter variants, with initial ego speeds
$1.0$, $1.5$, $2.2$, and $3.0\,\mathrm{m\,s^{-1}}$.
The geometry of the environments is a typical combination of straight and curved roads. The straight roads have length $20\,\mathrm m$, while the curved roads follow an S-curve with length parameter $20\,\mathrm m$ and amplitude $3\,\mathrm m$.

We take snapshots at control steps 0 and 8 of each of the 24 source encounters in order to get 48 total contexts. Ten evaluation streams per context give 480 attempts.
The two snapshots from an encounter are related; the reported availability
fractions are descriptive counts over this fixed benchmark. The fixed context includes ego and obstacle states, goal, reference speed,
reference path, obstacle-mode histories, prediction models, deterministic
anchors, and projection normals. The prediction horizon has 20 stages spaced
by $0.2\,\mathrm s$; source encounters use a $0.05\,\mathrm s$ control period.

\subsection{Fixed-normal constraints and the envelope}
\label{app:experiment_constraint_catalogue}

For each context $c$, the catalogue contains one coordinate for every
obstacle, trajectory model, and prediction stage $t=1,\ldots,20$.
Its dimension $G_c$ is therefore 20, 60, or 180 for the three configurations.
The normal $a_{c,g}$ points from the corresponding zero-noise obstacle
prediction toward a deterministic ego anchor. If this displacement has
norm at most $10^{-9}\,\mathrm m$, the previous normal is retained;
if none is available, the reference-path normal is used.
Normals and catalogue order are independent of the evaluation draws.

For an observation containing the trajectory model of collision-coordinate $g$, write its collision
half-space as
\begin{equation}
 a_{c,g}^{\top}p_t(\theta)\ge b_{c,g}(\delta),
 \qquad
 b_{c,g}(\delta)=\rho+a_{c,g}^{\top}p_{o,t}(\delta),
 \label{eq:mpc_group_lhs}
\end{equation} where $\rho$ is a specific radius offset. On the supported
numerical domain, a conservatively inflated score
$H_{c,g}(\delta)=\widetilde b_{c,g}(\delta)\ge b_{c,g}(\delta)$
is computed as described below. The solver receives one row per
coordinate of $E_N^{(c)}=\bigvee_i H_c(\delta_i)$, in fixed catalogue order.
The complete solver input
is unchanged by exact grouping
. 
\subsection{Numerical validity and nonrisk gates}
\label{app:numerical_gate}

The risk-transfer relationship in \eqref{eq:main_mpc_transfer} requires two conditions: the scenario scores must conservatively bound the primitive collision constraints, and the returned MPC proposal must satisfy the resulting sampled envelope. Since both conditions are evaluated numerically, we distinguish the exact mathematical quantities from their floating-point approximations.

\paragraph{Conservative collision offsets.}

Recall \eqref{eq:mpc_group_lhs}. Because the obstacle prediction model is linear in its Gaussian noise and the projection normal $a_{c,g}$ is fixed before sampling, this offset is an affine function of the noise. Writing $z\in\mathbb R^{40}$ for the response vector of the corresponding obstacle, we have
\begin{equation}
\label{eq:affine_collision_offset}
b_{c,g}(z)
=
\mu_{c,g}+L_{c,g}z,
\end{equation}
where $\mu_{c,g}\in\mathbb R$ is the deterministic offset obtained with zero noise and $L_{c,g}\in\mathbb R^{1\times40}$ describes the influence of the Gaussian noise on the projected obstacle position. Both coefficients are determined by the frozen prediction model and projection normal. In numerical evaluation, the computed offset $\widehat b_{c,g}(z)$ may differ from the exact expression \eqref{eq:affine_collision_offset} because of floating-point rounding and coefficient-construction errors.

To preserve the required conservative inequality, the implementation associates each coordinate with a nonnegative error bound $e_{c,g}^{\mathrm{off}}$ satisfying $b_{c,g}(z)
\le
\widehat b_{c,g}(z)
+
e_{c,g}^{\mathrm{off}}$ throughout the supported numerical domain. The inflated score $\widetilde b_{c,g}(z)$ is obtained by rounding this upper bound outward, so that
\begin{equation}
\label{eq:conservative_offset_condition}
b_{c,g}(z)
\le
\widetilde b_{c,g}(z).
\end{equation} 

\paragraph{Supported numerical domain.}

The error bounds used in \eqref{eq:conservative_offset_condition} are established only for a specified range of numerical inputs. In particular, we require 
\[
\|z\|_\infty\le12
\]
for each obstacle, together with the implementation's prescribed bounds on states, model coefficients, projection normals, and floating-point evaluation. If an observation violates any supported-domain condition, its complete score vector is assigned $+\infty$ in every coordinate. Consequently, any evaluation bank containing such an observation is rejected by the common feasibility gate.

\paragraph{Verification of the MPC proposal.}

After constructing the sampled envelope
\[
E_N^{(c)}
=
\bigvee_{i=1}^{N}H_c(\delta_i),
\]
the MPC solver returns a candidate decision $\theta_N$ and its predicted ego trajectory. For each group with a finite envelope coordinate, define the exact constraint residual
\begin{equation}
\label{eq:mpc_constraint_residual}
r_{c,g}(\theta_N)
:=
a_{c,g}^{\top}p_{t(g)}(\theta_N)
-
E_{N,g}^{(c)}.
\end{equation} A nonnegative residual means that the corresponding sampled envelope constraint is satisfied. Because the residual is also evaluated numerically, the implementation computes a conservative lower bound $\underline r_{c,g}$ satisfying $\underline r_{c,g}
\le r_{c,g}(\theta_N)$. The lower bound accounts for arithmetic errors in evaluating the projected ego position and subtracting the envelope offset. The returned proposal passes the collision-constraint verification only if
\begin{equation}
\label{eq:verified_mpc_feasibility}
\underline r_{c,g}\ge0
\qquad
\text{for every active group }g.
\end{equation} Provided the numerical error bounds are valid, this check establishes
\[
a_{c,g}^{\top}p_{t(g)}(\theta_N)
\ge E_{N,g}^{(c)}
\]
for every active envelope coordinate, independently of the solver's convergence tolerance.

\paragraph{Common feasibility gate.}

We denote by $\mathsf{Gate}_{N,c}$ the indicator that all required numerical and deterministic checks pass. These checks include successful MPC optimization without fallback, convergence of the final quadratic-programming subproblem, successful completion of the implementation's support-processing iterations, no computational support-cap termination, and strict satisfaction of the configured state, input, and road constraints. Gate acceptance additionally requires that the sampled envelope contain no $+\infty$ coordinates and that every residual lower bound satisfy \eqref{eq:verified_mpc_feasibility}. A failed check causes rejection without modifying the primitive observations.

The gate is evaluated before statistical certificate acceptance and is identical for all certificate-accounting methods. In particular, changing the block size $k$ or evaluating primitive Wait-and-Judge certification does not change the sampled envelope, MPC proposal, or numerical feasibility checks. On the event $\{\mathsf{Gate}_{N,c}=1\}$, the conservative offset condition \eqref{eq:conservative_offset_condition} and the verified envelope constraints imply
\[
V_{P_c}(\theta_N)
\le W_{P_c}(E_N^{(c)}),
\]
as stated in \eqref{eq:main_mpc_transfer}. Thus, the common numerical gate supplies the feasibility condition needed to transfer the envelope certificates to modeled primitive MPC constraint violation.

\subsection{Witnesses and certificate computation}
\label{app:experiment_cert_alg}

For every active coordinate, the canonical witness retains the least
primitive index attaining its stored maximum.
For consecutive blocks of size $k$, these indices are mapped to
$1+\lfloor(i-1)/k\rfloor$ and deduplicated to obtain $r_k$.
Indices in the saved implementation are zero-based, giving the equivalent
map $i\mapsto\lfloor i/k\rfloor$.
Ties use exact equality of stored offsets, with no tolerance.
This witness reproduces the envelope though it is not necessarily minimal and it is not an
estimate of the nonlinear optimizer's support constraints.

Elementary acceptance requires the common gate, $r_k<M_k=N/k$, and
$\Psi_k(N,r_k)\le\beta$. At the experimental parameters, offline acceptance
is checked by the exact integer comparison
\[
 100M_k\binom{M_k}{r_k}19^{N-kr_k}\le20^{N-kr_k}.
\]
The recorded decisions are checked against this comparison, including
the support boundary. Primitive WJ uses $N$ and $r_{\mathrm{prim}}$,
independently of $k$. Rather than numerically approximating the root in
\eqref{eq:wj_root}, its offline implementation tests, for $r<N$,
\[
 \frac{\beta}{N}\sum_{m=r}^{N-1}\binom mr q^{m-r}
 \ge \binom Nr q^{N-r},
 \qquad q=19/20.
\]
This is equivalent to $\varepsilon_N^{\mathrm{WJ}}(r)\le\epsilon$;
the finite sum is evaluated through an exact integer binomial-tail identity
using rational $\epsilon=1/20$ and $\beta=1/100$.
Equality accepts, while $r\ge N$ rejects.

\subsection{Additional experimental results}
\label{app:experiment_primary}

In addition to the results presented in the main text, we offer a few supplementary details on our simulation. For our primary analysis we reuse the
$N=3816$ primitive banks from the
archived $k=12$ experiment. All alternative certificate assessments
are performed retrospectively on the
same saved banks and MPC proposals. An earlier experiment using
$N=1351$ primitive observations is
excluded from the paired comparison
because it uses a different
information budget.

\noindent The accounting-only sweep evaluates
\[
k\in\{1,2,3,4,6,8,12\},
\]
yielding respective accepted counts,
each out of 480 attempts, of
\[
318,\ 350,\ 405,\ 449,\
471,\ 471,\ 471.
\]

\noindent As mentioned before, we see that primitive WJ accepts the same
471 individual attempts as
$k=12$ elementary certification.
All nine common-gate rejections
fail strict deterministic admissibility. The corresponding elementary
support-cardinality ceilings are
\[
31,\ 36,\ 40,\ 43,\
48,\ 52,\ 58.
\] Observed primitive witness
cardinalities range from 2 to 49.

\begin{table}[ht]
\centering
\caption{Fresh-bank certification counts, each out of 480 attempts.
The final column counts common nonrisk rejections.}
\label{tab:appendix_fresh_budget}
\begin{tabular}{r|rrr|r}
\hline
$N$ & SS, $k=1$ & SS, $k=12$ & Primitive WJ & Rejected\\
\hline
1536 & 218 & 318 & 476 & 4\\
1920 & 297 & 319 & 468 & 12\\
2304 & 317 & 386 & 473 & 7\\
2688 & 312 & 451 & 468 & 12\\
3072 & 315 & 471 & 472 & 8\\
3456 & 317 & 473 & 473 & 7\\
3816 & 323 & 475 & 475 & 5\\
4320 & 339 & 467 & 467 & 13\\
4800 & 390 & 473 & 473 & 7\\
5760 & 472 & 476 & 476 & 4\\
\hline
\end{tabular}
\end{table}

The fresh-budget study reported in Table~\ref{tab:appendix_fresh_budget} evaluates
480 independently generated primitive
banks at each of ten tested budgets.
Within a budget, all compared
certificates use the same bank,
envelope, numerical proposal, and
common feasibility gate. Across the ten budgets, 79 attempts
fail the common gate. Primitive WJ
accepts every remaining attempt
in the reported benchmark.

The first tested budgets attaining
at least $95\%$ certification
availability are $5760$, $3072$,
and $1536$ for singleton elementary,
$k=12$ elementary, and primitive WJ
certification, respectively. These are empirical crossings on
the tested budget grid, not
distribution-free minimum sample
requirements. Different budgets
use newly generated banks, and their
acceptance counts need not increase
monotonically. 

\end{document}